\documentclass[sigconf]{acmart}
\usepackage{accsupp}
\usepackage{url}
\usepackage{multirow}
\usepackage{CJK}

\copyrightyear{2026}
\acmYear{2026}
\setcopyright{cc}
\setcctype{by-nc-nd}
\acmConference[CHI '26]{Proceedings of the 2026 CHI Conference on Human Factors in Computing Systems}{April 13--17, 2026}{Barcelona, Spain}
\acmBooktitle{Proceedings of the 2026 CHI Conference on Human Factors in Computing Systems (CHI '26), April 13--17, 2026, Barcelona, Spain}
\acmPrice{}
\acmDOI{10.1145/3772318.3790464}
\acmISBN{979-8-4007-2278-3/2026/04}

\begin{document}

\title[Access in the Shadow of Ableism] {Access in the Shadow of Ableism: An Autoethnography of a Blind Student's Higher Education Experience in China}


\author{Weijun Zhang}
\authornote{The two authors contributed equally to this work. The authorship order is interchangeable. We listed Zhang first here to acknowledge that his lived experiences form the foundation of this work.}
\orcid{0009-0007-3344-8855}
\affiliation{\institution{Syracuse University}\city{Syracuse}\state{New York}\country{USA}}
\email{wzhan124@syr.edu}

\author{Xinru Tang}
\authornotemark[1]
\orcid{0000-0001-6426-1363}
\affiliation{\institution{University of California, Irvine}\city{Irvine}\state{California}\country{USA}}
\email{xinrut1@uci.edu}


\begin{abstract}
The HCI research community has witnessed a growing body of research on accessibility and disability driven by efforts to improve access. Yet, the concept of access reveals its limitations when examined within broader ableist structures. Drawing on an autoethnographic method, this study shares the co-first author Zhang's experiences at two higher-education institutions in China, including a specialized program exclusively for blind and low-vision students and a mainstream university where he was the first blind student admitted. Our analysis revealed tensions around access in both institutions: they either marginalized blind students within society at large or imposed pressures to conform to sighted norms. Both institutions were further constrained by systemic issues, including limited accessible resources, pervasive ableist cultures, and the lack of formalized policies. In response to these tensions, we conceptualize access as a contradictory construct and argue for understanding accessibility as an ongoing, exploratory practice within ableist structures.
\end{abstract}



\maketitle
\section{Introduction}
\begin{quote}
    \textit{Access didn't eliminate ableism; it enabled ableism to bare its teeth.} -- \textit{Crip Negativity, J. Logan Smilges~\cite{smilges2023crip}} 
\end{quote} 
Accessibility and disability have increasingly been key themes in HCI research~\cite{mack2021we}. Oftentimes, this body of work is driven by a language of access, as seen in the growing amount of work on assistive technologies or work that seeks to make technologies more accessible to people with disabilities (PWD)\footnote{We use both people-first (people with disabilities) and identity-first (disabled people) language to recognize diverse naming preferences within disabled communities.}~\cite{mack2021we}. 
As J. Logan Smilges observes,
\begin{quote}
    \textit{``The field of disability studies and the wider terrain of disability activism often frame ableism as a collection of access problems. If a person can't enter a room because there is no ramp, if they can't participate in a conversation because there is no sign language interpreter, or if they can't understand a contract because it is too full of unnecessary jargon, then that room, conversation, and contract are inaccessible. Inaccessibility occurs when disabled people and our needs are intentionally dismissed or accidentally overlooked. Either way, ableism is at work, barring disabled people from full participation.''}~\cite{smilges2023crip}
\end{quote}
Nevertheless, emerging critiques have highlighted the limits of such narratives, revealing that ableism extends beyond what the concept of access can fully capture, and at times may unintentionally reproduce ableism~\cite{smilges2023crip, shew2023against, williams2023counterventions, williams2019prefigurative}. An increasing body of scholarship suggests that access is often embedded within larger ableist structures or systems that pressure PWD to conform and be included~\cite{smilges2023crip, shew2023against}. These critical perspectives have increasingly informed accessible computing research, motivating efforts toward more responsible design for disabled communities~\cite{williams2023counterventions, williams2019prefigurative, mankoff2010disability}. However, many gaps remain. For instance, most critiques have been made within what researchers call WEIRD (Western, Educated, Industrialized, Rich, and Democratic) societies~\cite{henrich2010most}, with far less attention paid to other cultural contexts. Additionally, substantial work remains to be done to understand how access unfolds within specific institutions.

This study contributes to the ongoing dialogue on access through a case study of a blind student’s experience in Chinese higher education. Drawing on an autoethnographic method~\cite{ellis2011autoethnography}, we report Zhang, the blind co-first author's experiences at two higher education institutions in China that employ distinct approaches to teaching blind and low vision (BLV)\footnote{The definition of blindness is inherently political and often personal. We use ``blind and low-vision'' to broadly refer to visual disabilities. We use ``blind'' to refer specifically to people who require accommodations such as screen readers or Braille, rather than relying on residual visions.} students. One institution is a specialized program exclusively designed for BLV students, while the other is a mainstream university primarily attended by sighted students. Through a critical reflection over Zhang's experiences, we reveal the dilemmas inherent in access when it is positioned in broader institutional and socio-cultural structures. In the context of special education, efforts to provide access often inadvertently marginalize students in the broader society. Conversely, negotiating access within a sighted-centric institution entails pressures to conform to sighted norms. Furthermore, access in both institutions was constrained by systemic issues, including limited accessible resources, ableist cultures, and the lack of established accessibility policies. Consequently, a blind student must navigate ableist realities and adapt to various systems to access future opportunities. This comparison led us to reflect on access as a lifelong journey and to acknowledge it as a contradictory concept shaped by broader ableist structures.

This study makes two major contributions to the HCI and accessible computing literature. First, we provide a first-hand understanding of higher education accessibility in China from a blind student's perspective. This understanding is valuable given the small number of blind students who pursue higher education in China, especially at the graduate level~\cite{higher-edu-blind-china, gaokao-blind-graduate, hu2022opportunities}. Zhang  (the blind co-first author) was the first blind student to take the admission test of his graduate program and be accepted. Second, our analysis reveals inherent tensions involved in seeking access. Rather than framing access as solutions to ableism, we argue that it should be understood as an ongoing, exploratory practice that unfolds within ableist structures. We conclude with implications for institutions, practitioners, activists, and future research to better support disabled students in navigating pervasive ableist systems.

\paragraph{Author Contributions}
This work is a collaborative effort between the two authors. Zhang's lived experiences form the foundation of this work. He curated the personal narratives used in this study and engaged with discussions throughout to help deepen the analysis. Tang was responsible for situating these experiences within the literature, interpreting the segments Zhang shared from an academic angle, developing the theoretical framing, and writing the paper.

\section{Background}
We provide a brief overview of the socio-cultural and legal context of disability in China to establish the necessary background for understanding our study.

\subsection{Disability in Chinese Culture and Legal Systems}
\label{section::culture-law}
The implementation of inclusive education in China has been strongly shaped by historical emphasis on elitism in traditional East Asian culture, which prioritizes standardized curricula, exams, and competitive achievements~\cite{huang2005elitism, deng2012reforms}. China, Japan, and South Korea have historically depended almost exclusively on standardized examinations to determine access to higher education and government employment~\cite{xie2024noncompliant}. Besides, disability has historically, and continues to be, perceived through medical and charitable frameworks in East Asia, often being framed as an individual deficit and personal or family responsibility~\cite{qu2020understanding, kim2025disclosing}. These cultural traditions have shaped societal attitudes toward disability and inclusive education more broadly. For example, the most recent government plan continues to prioritize special education and the expansion of special schools as the primary strategy for educating disabled students~\cite{special-edu-policy-gov}.

Furthermore, China's inclusion policy is still in an early stage of development and currently lacks clear, systematic, and coherent implementation strategies~\cite{xie2024noncompliant, qu2022structural}. While there is evidence of improved educational conditions for PWD, especially following China’s ratification of the United Nations Convention on the Rights of Persons with Disabilities (CRPD), scholars in law and policy have argued that inclusive education policies remain largely vague and often fail to fully comply with the CRPD~\cite{xie2024noncompliant, qu2022structural}. Within the Chinese legal framework, \textit{people with disabilities} are consistently defined as individuals whose impairments result in functional limitations rather than following the social model under CRPD~\cite{law-disability-china}. Besides, although legislation prohibits discrimination against PWD, Chinese law has not yet defined \textit{discrimination} or provided a clear concept of \textit{reasonable accommodation}. Currently, the sole reference to \textit{reasonable accommodations} appears in Article 52 of the 2017 Chinese Education Regulation, which is limited to \textit{``when persons with disabilities take the national examinations.''}~\cite{rc-policy} Consequently, students’ rights to reasonable accommodations outside the context of national examinations are not legally guaranteed. These policy gaps have left disabled people and their families facing challenges in addressing discrimination.

\subsection{Higher Education for Blind Students in China}
In China, attendance in higher education among students with disabilities remains low. According to the Chinese government, only slightly more than 50,000 PWD\footnote{The number refers to individuals officially registered as disabled under China’s classification framework~\cite{china-disability-classification}.} attended higher education between 2016 and 2020~\cite{higher-edu-disability-china}. In contrast, the number of non-disabled students enrolled in higher education was 47.63 million in 2023~\cite{higher-edu-china}. The number of BLV students is likely smaller. A 2020 news report noted that only around 200 BLV students enrolled in undergraduate programs each year~\cite{higher-edu-blind-china}\footnote{This number represents BLV students who entered higher education institutions through ``specialized exams and admissions.'' Details of this system are provided in the following paragraph. Accurately determining higher education attendance among BLV students is difficult not only because of a lack of comprehensive census~\cite{hu2022opportunities}, but also because many disabled students might choose not to disclose their disabilities to avoid discrimination.}.

In China, the dominant model to provide accessible education for disabled students continues to operate within the special education system. At the K–12 level, China still emphasizes the construction of specialized schools~\cite{xie2024noncompliant} and expects students with disabilities to adapt to mainstream classrooms rather than providing reasonable accommodations~\cite{qu2020understanding}. In higher education, the commonly used model, known as ``specialized exams and admissions'' (\begin{CJK}{UTF8}{gbsn}单考单招\end{CJK}), was introduced in the 1980s. In this model, BLV students sit for a small number of specialized exams offered by a few universities that offer specialized education, rather than the national college entrance examination (\textit{``gaokao.''})~\cite{gaokao-blind} Their choice of majors is also highly restricted, typically to fields such as acupuncture, remedial massage, and music~\cite{ma2020hero}. Remedial massage has often been regarded as the default major for blind students due to the strong historical emphasis on vocational education within blind education~\cite{li2022protection}.

In 2017, the Chinese Ministry of Education established policies that ensure that disabled students can request reasonable accommodations on \textit{gaokao}, including Braille exam papers and extended time~\cite{rc-policy}. However, very few BLV students opt for the \textit{gaokao}, as they must decide between it and the \textit{specialized exams and admissions}, with no option to take both. Most BLV students who need accommodations still choose to take the latter because the quality of K-12 education in the special schools they attend is much lower than mainstream schools~\cite{ma2020hero, hu2022opportunities}. Statistics from 2014 to 2020 showed that out of the tens of millions of \textit{gaokao} candidates, fewer than ten students requested Braille exam papers each year~\cite{hu2022opportunities}. In 2021, this number rose slightly to 11~\cite{hu2022opportunities}. As a result, most mainstream higher education institutions in China remain largely unaware of and inexperienced in supporting BLV students.

\section{Related Work}
Our study is informed by the growing scholarship of critical disability and access studies, research on higher education accessibility, and autoethnography studies in HCI and accessibility.

\subsection{Critical Disability and Access Studies}
Critical disability and access studies constitute an expansive field that seeks to challenge assumptions surrounding disability and access, viewing both topics as subjects subject to ongoing ``debate and dissent.''~\cite{kafer2013feminist, hamraie2019crip} For instance, Hamraie et al. coined the term ``crip technoscience,''~\cite{hamraie2019crip} shifting the attention of access design from technological solutions to knowledge generated by disabled individuals and communities, and politics surrouding assistive technologies. Scholars note that, in some cases, technologies might even reinforce ableist assumptions as they operate in broader ableist structures~\cite{shew2023against, kafer2013feminist}.

Mankoff et al. introduced critical disability studies as a key area of inquiry for accessible computing in 2010~\cite{mankoff2010disability}. Since then, the accessibility research community has seen an increasing number of critical reflections on access and disability~\cite{hofmann2020living, bennett2018interdependence, bennett2019promise, williams2023counterventions, ymous2020just, spiel2019agency, sum2024challenging, tang2024accessibility}, along with greater engagement with critical disability studies in empirical studies~\cite{alharbi2024misfitting, hsueh2023cripping, angelini2025speculating, brule2019negotiating}. This growing body of work has prompted deeper reflection on the models of disability embedded in assistive technology design, with an increasing number of studies incorporating alternative perspectives, such as social, cultural, and relational models of disability, to inform research and design~\cite{tang2026disability, tang2026reimagining, tang2023community, tran2026toward, angelini2025speculating, williams2023counterventions, alharbi2024misfitting, mcdonnell2024envisioning, hsueh2023cripping}. Furthermore, research has cautioned against harmful patterns in accessibility design and practice, advocating for disability involvement throughout the research and design process~\cite{williams2023counterventions, bennett2018interdependence, gamage2023blind, mack2021we, garg2025s, tang2026disability}. 

The broader socio-political context that shapes access is surfaced through these ongoing reflections~\cite{williams2023counterventions, williams2019prefigurative, das2019doesn, cha2025dilemma, saha2021urban, apara2025, ly2025accessibility}. For example, research has highlighted the tension between medical systems and the lived experiences of disabled people~\cite{williams2023counterventions, tichenor2019stuttering}, institutional interests versus individual needs~\cite{cha2025dilemma, apara2025}, and other competing priorities that influence accessibility in practice~\cite{saha2021urban, apara2025}. These interconnected complexities reveal ableism as a systemic issue that requires thorough examination and careful considerations. However, most critiques have been conducted in the U.S.~\cite{mack2023maintaining, shinohara2020access, shinohara2021burden, tamjeed2021understanding, williams2023counterventions, williams2019prefigurative}, leaving limited insight into how disability and access vary in other cultural and political contexts. Although accessibility research conducted in contexts known as Global South and non-WEIRD regions is growing, much of this work has focused primarily on the empirical exploration of individuals' adoption or use of assistive technologies~\cite{pal2017agency, kameswaran2018we, chandra2021piracy, kameswaran2023advocacy, barbareschi2021difference}, with less emphasis on socio-political and institutional influences. Other studies have concentrated on accessibility policies~\cite{nourian2022digital} or infrastructure~\cite{pal2016accessibility}, but still lack understanding of disabled people's personal experiences. Our work seeks to extend critical scholarship on disability and access through an autoethnographic account of a blind student's experiences in higher education in China.

\subsection{Higher Education Accessibility}
As a vital public service, higher education institutions have been central to studies in accessible computing~\cite{mack2023maintaining, shinohara2020access, tamjeed2021understanding, shinohara2021burden, yildiz2023institutional, ly2025accessibility}. Research has revealed a variety of accessibility issues within higher education systems, covering physical spaces~\cite{dolmage2017academic}, classroom setups~\cite{price2011mad}, course materials~\cite{shinohara2020access}, and technological tools~\cite{shinohara2020access, shinohara2021burden}. Studies also emphasize the need for clear communication and collaboration among key stakeholders, as a lack of shared understanding often hinders effective support~\cite{yildiz2023institutional, mack2023maintaining}. Many staff members or faculties may still attribute the `problem' of disability to disabled students themselves ~\cite{yildiz2023institutional}. Ultimately, the effectiveness of accessibility efforts often depends on professors' willingness~\cite{tamjeed2021understanding}. Even when instructors are willing, they may still face challenges in implementing accommodations or supporting disabled students due to a lack of experience~\cite{dunn2012assisting}. This wide range of issues has made higher education accessibility a complex network of interconnected stakeholders~\cite{mack2023maintaining, ly2025accessibility}.

Disabled students often take on significant labor to meet their access needs~\cite{jain2020navigating, shinohara2021burden}, such as spending considerable time and effort requesting access services (e.g., interpreters), incorporating accommodations into their routines (e.g., installing and learning new software), and developing workarounds to address pervasive inaccessibility in technological tools~\cite{shinohara2021burden}. Additionally, seeking accommodations can be complicated by concerns over discrimination, leading to additional emotional cost~\cite{jain2020navigating}. Despite extensive research, much of the existing work has focused on contexts where disability services offices play a central role in providing accommodations, particularly in the U.S.~\cite{mack2023maintaining, shinohara2020access, tamjeed2021understanding, shinohara2021burden}. Our study seeks to expand this literature with a case in China, where formal accommodation services remain limited~\cite{hu2022opportunities}.

\subsection{Autoethnography in HCI and Accessibility}
Autoethnography is a qualitative research method in which researchers take on the dual role of participant and analyst within an ethnographic study~\cite{adams2021introduction}. It involves self-observation and critical reflection to explore personal experiences within broader cultural, political, and social contexts~\cite{adams2021introduction}. Autoethnography holds unique value by providing research communities with detailed, firsthand insights and a deep understanding of individuals' lived experiences~\cite{kaltenhauser2024playing}. The method has gained increasing recognition in HCI and accessibility research over the past decade~\cite{kaltenhauser2024playing}, offering an intimate understanding of disability~\cite{jain2020navigating, wu2024finding, cassidy2024dude, le2024human, hofmann2020living, williams2019prefigurative, aishwarya2022performing}. Additionally, autoethnography has been recognized as an accessible research method as many traditional ethnographic field sites present accessibility issues~\cite{patrycja2012autoethnography}.

In accessible computing, autoethnographic studies have provided valuable insights into broader contexts shaping accessibility and informed the design of assistive technologies~\cite{jain2020navigating, wu2024finding, cassidy2024dude, le2024human, hofmann2020living, williams2019prefigurative, aishwarya2022performing}. A key study relevant to our research is by Jain et al., who conducted a trio-ethnography reflecting on their experiences with graduate education in the U.S.~\cite{jain2020navigating} They explored how factors such as self-image, relationships, technology, and institutions impact their daily access. Their findings highlight the complexities and tensions within graduate school, offering a nuanced understanding of the social-cultural negotiation process and the emotional cost involved in gaining access. Similarly, Mack et al. combined autoethnographic methods and interviews, uncovering the need for effective collaboration among multiple stakeholders to ensure accessibility in the U.S. higher education system~\cite{mack2023maintaining}. Both studies have provided critical insights into design opportunities for assistive technologies, including collaborative systems for accommodations and proactive, and customizable assistive tools~\cite{mack2023maintaining, jain2020navigating}. Building on the strengths of autoethnography, our study provides an intimate insight into blind education within higher education in China.

\section{Methods}
\subsection{Data Collection and Analysis}
Our methods combined the reflexive approach outlined by Hoffman et al.~\cite{hofmann2020living}, and auto-ethnographic practices commonly adopted in HCI and accessibility research~\cite{jain2020navigating, duncan2004autoethnography}. Specifically, our study consists of two main components: (1) the collection of personal narratives and reflections, and (2) a reflexive thematic analysis~\cite{braun2021can}. In December 2023, Zhang developed a retrospective account of his higher education experiences as part of a course project. In line with practices adopted in previous autoethnographies in HCI and accessibility~\cite{jain2020navigating}, he thoroughly recalled his experiences at two higher education institutions he attended in China, including the admissions process, the ways technology was adopted and used, and the accessibility issues he faced. This reflection resulted in over 20 vignettes, which were categorized under two institutions and organized around themes covering instructional design, physical environments, digital accessibility, and exam accommodations.

Tang joined the study later to bring a critical perspective and her expertise in HCI and accessible computing to the analysis and writing. Together, we re-analyzed the story fragments generated in the first stage and inductively developed a set of themes to probe how access was practiced in the two institutes. We adopted reflexive thematic analysis for our analysis, which entails iterative and ongoing theme development along with data collection based on patterns of shared meaning among the data~\cite{braun2021can}. Building on insights from the first stage, we posed deeper questions to engage more thoroughly with the data, such as: \textit{What do we expect access to do? What are the benefits of securing access? What factors shapes and influences the process of gaining access?} During this phase, we also reviewed Zhang's social media accounts to aid in memory recall. Additional memories emerged and were integrated into the narrative set. This iterative process continued until we reached thematic saturation.

\subsection{Biography}
We provide Zhang's background to contextualize this autoethnography within the Chinese context. Zhang became blind at a young age due to glaucoma. Like most blind students in the country, he attended special schools for his K-12 education as this is the major model in China providing accommodations to disabled students~\cite{ma2020hero, qu2022structural}. However, while national policies direct most BLV students in China toward vocational instead of post-secondary education~\cite{hu2022opportunities}, he chose to attend an academically focused high school to prepare for higher education.

Our study reports Zhang's experiences at two institutions he attended for his undergraduate and graduate study. We refer to them as University A and University B throughout the rest of this paper. University A represents the predominant special education model to provide accessible higher education to BLV students in China. Although the institution admits both disabled and non-disabled students, it restricts students with visual and/or hearing disabilities, to a limited set of majors within its College of Special Education. Additionally, these students are required to follow a fixed curriculum with no opportunity to choose courses.

University B is a mainstream institution specializing in foreign language studies, where our blind author completed his graduate education. Notably, attending graduate school remains uncommon among blind students in China due to a longstanding lack of accommodations and policy support. Zhang was the first blind student to take the university's admission exam, as well as the first in the province where he took the exam. He was also the first blind student in the country to apply for a major in translation and interpreting. Finally, he became the first and only blind student admitted to the university when he studied there.

\subsection{Positionality}
We drew on both our personal experiences and academic perspectives in shaping this study. Both authors were raised in China and currently study at academic institutions in the U.S. Zhang identifies blind and studied in China prior to pursuing his second Master's degree in the U.S. His understanding of disability and accessibility has been shaped by decades of living with a visual disability, and studies and work in disability-related fields. Tang identifies sighted. She brought an HCI and critical perspective to this study informed by her research with diverse disabled populations. While the global disability research and activism, especially in the U.S., influenced our understanding of disability justice, we weighed in the realities that disabled people in China face. Additionally, we are mindful of critiques regarding the role of critical disability studies in accessible computing research~\cite{kirkham2021disability}. Our goal in writing this autoethnography is not to provide a comprehensive account of higher education accessibility in China or to offer prescriptive solutions, but to encourage critical reflection on how accessibility is and could be practiced.

\subsection{Reflection on Methods}
As with all research methods, autoethnography has its unique values and limitations~\cite{mendez2013autoethnography}. Notably, it is often valued for its explicit subjective positioning~\cite{foster2006extending}, which contrasts with approaches that aim for more objective accounts. Here, we offer a critical reflection on our methods and invite readers to engage thoughtfully with our work. First, our data collection has primarily relied on our blind author's self-account. This approach offers a level of intimacy and personal account rarely achievable through other methods. However, the reliance on one's memory may introduce bias and affect the accuracy of the account. For example, although inaccessibility in technology was a common experience, Zhang could not recall all the specific issues he encountered when using specific technologies. We encourage readers to view our work as a valuable case of a blind student's educational experiences from a personal perspective, rather than a generalizable account. Still, we encourage readers to be attentive to the lessons our study can offer for other contexts, since many of the broader influences are widely relevant such as pervasive inaccessibility disabled people encounter in everday life~\cite{tang2025every}. Second, autoethnography is often deeply emotionally driven~\cite{mendez2013autoethnography}. Our analysis and writing might have used strong or emotionally charged language, especially when we tried to employ an idealized standard to provoke our radical reflection on access. While sometimes criticized as self-indulgent~\cite{mendez2013autoethnography}, these emotional accounts serve a valuable lens for understanding the affective dimensions central to the lived experiences of disability. We encourage future research to build on our work by incorporating other methods, such as interviews and by involving multiple ethnographers to bring other perspectives.

\section{Findings}
Our analysis reveals tensions inherent in creating access. While both institutions Zhang attended worked to make education more accessible to students, their efforts were constrained by broader systemic structures and conceptions of disability.

\subsection{University A: A Special Program for Blind Students}
Zhang attended the College of Special Education at University A for his undergraduate study, which placed students with visual and/or hearing disabilities in a specialized program. This exclusive environment made it easier to provide accommodations for disabled students. However, we hesitated to describe University A as accessible because it was shaped and constrained by broader ableist systems and conceptions. 

\subsubsection{The Dilemma Under Special Education}
A major tension surfaced in our analysis is a dilemma rooted in the long-standing history of special education for disabled students in China. This is evident in the limited academic options available to Zhang.
\begin{quote}
``\textit{When I graduated from high school in 2013, the only options I had for higher education were the admission exams offered by two universities that admitted blind students. What's more, I could only choose a major in acupuncture and remedial massage.''}
\end{quote}
The restrictions on major choice stem from long-standing policies in China that encourage remedial massage as a means of enhancing employment opportunities for blind people~\cite{li2022protection}. While some believe this policy has improved the employability of blind people, over the decades, massage has become the primary, and often the only-viable career path available to blind people in China \cite{li2022protection}.

The mismatch between his learning interests and the available majors led to a decline in Zhang's academic performance. This drop in performance made his high school teacher question his commitment to learning.
\begin{quote}
``\textit{I failed a course in my first year of university because I didn't like the major...My high school teacher wondered why, as a top student in high school, I was failing university courses and whether I was making enough effort.}''
\end{quote}
The quote above shows that while limited major options are the root cause of Zhang's declining performance, people might place the responsibility on disabled students, creating pressure to conform.

On the positive side, the specialized system at University A created an environment where BLV students could be the `majority'. A clear example of this benefit is that, despite the growing popularity of PowerPoint at the time, the college chose not to adopt it.
\begin{quote}
``\textit{Throughout my undergraduate studies, we almost never used PowerPoint slides in class. To be honest, this was beneficial for BLV people, as PowerPoint is a product of visual culture. It is inefficient and unfriendly for screen reader users.}'' 
\end{quote}
Beyond the classroom, BLV students were also assigned to shared dormitories, creating an environment where they could exchange information when technology was inaccessible. For instance, the class created a QQ chat group to communicate routine information, such as class meetings. While the application was not accessible to some students, their roommates could help relay the information. These examples highlight the tensions inherent in learning within a special education system. On one hand, the institute made efforts to make education accessible to BLV students. However, these efforts could inadvertently reinforce marginalization within a broader ableist and visually oriented society.

\subsubsection{Access Hindered by Systemic Inaccessibility}
Efforts to provide access at University A were further constrained by a systemic shortage of accessible resources. A typical example is Braille books: not all textbooks had Braille versions, and even when they did, they were updated far less frequently than the standard versions. As a result, students who preferred to use Braille had to develop ad-hoc solutions if instructors needed the latest versions of textbooks for teaching.
\begin{quote}
    \textit{``Blind students might have to ask former students for digital copies, while some created Braille versions of essential content themselves and shared with others. In some cases, instructors might have to continue to use older versions considering the number of blind students.''}
\end{quote}
The lack of accessible resources was further exacerbated by the institute's limited budget for acquiring up-to-date tools. For example, the computers at the school only had an outdated free version of a screen reader, released eight years ago, and many shortcut keys and interaction methods had changed in subsequent updates.

Another systemic issue is the lack of training among instructors. At the time of Zhang's study, none of the instructors could read or write Braille, which restricted students' options for learning through Braille.
\begin{quote}
    \textit{``Blind students had to write essays on computers and relied on low-vision peers to print their assignments. Formatting articles is inaccessible to most students using screen readers, forcing them to rely on plain text versions or perform only the most basic typing in Word for their assignments. Although students were permitted to take final examinations in Braille, the assessment process relied on translation provided by students proficient in Braille.''}
\end{quote}
This quote illustrates how blind students had to adapt to inaccessible environments through self-developed workarounds, often at the cost of their learning experiences. While instructors and staff may strive to improve accessibility, these issues are often systemic and cannot be addressed by a single person or institution.

\subsubsection{Access Constrained in Ableist Cultures}
A deeper tension arises between the flexibility required for accessibility and the traditional culture of elitism in China. As mentioned in Section \ref{section::culture-law}, the education system in China places a strong emphasis on standardized tests, and this is also true for University A. Blind students were expected to follow the same curriculum, course content, assignments, and assessment methods as sighted students, under the assumption that applying identical standards ensured fairness. However, many course materials reflected sighted norms, and teaching practices rarely incorporated non-visual instructional approaches.
\begin{quote}
    ``\textit{When instructors encountered visual material, they often told students to memorize it or simply skipped those sections, assuming that blind students could not engage with the content.''}
\end{quote}
Additionally, as noted in \ref{section::culture-law}, Chinese culture often places the responsibility of navigating the ableist society on disabled individuals themselves. This is evident when instructors prioritized convenience in teaching over accommodating blind students' needs.
\begin{quote}
    ``\textit{During the practice lessons on massage techniques, some instructors found it inefficient to provide one-on-one instruction for fully blind students. As a result, blind students were often forced to learn from peers who have mastered the technique. This approach typically results in suboptimal learning outcomes.}''
\end{quote}
In the example above, the lack of individualized support and non-visual teaching methods shifted the burden of adaptation onto the students and ultimately limited the quality and depth of their learning experience.

Moreover, reflecting the prevailing medical-individual model of disability in China, when a student failed to meet pre-determined standards, the responsibility was often attributed to the student's abilities. For example, one instructor required students to complete a set number of warm-up exercises such as push-ups. These exercises were considered essential for preparing bodies to practice massage. However, some students could not perform some exercises because doing so posed a risk of worsening their eye conditions.
\begin{quote}
\textit{``When these students asked the instructor for alternative ways to complete the assessment, they received responses that invalidated their capabilities. The instructor told one student, ``considering your physical condition, you are not suited for studying massage. I suggest you quit the school immediately.''''}
\end{quote}
As this quote suggests, pervasive ableist socio-cultural systems continue to shape attitudes toward disabled students, which often constrain their learning experiences, and potential.

\subsection{University B: A Sighted-Centric Institution}
University B is a mainstream institution where Zhang was the first blind student to be admitted. As a result, sighted students not only formed the majority but also dominated the norms and practices of the institution -- what we refer to as a sighted-centric environment. While the university made efforts to provide accommodations to Zhang, we identified two tensions at both the institutional and socio-cultural levels: the lack of established policies and the pressure to conform to sighted norms. Although some opportunities for change were available, they remained ad hoc.

\subsubsection{Lack of Established Policies}
A key tension that emerged from our analysis is the lack of established policies and dedicated entities, such as a disability center or writing center, that could provide consistent accessibility support. As a result, obtaining access often requires coordination across multiple parties and depends heavily on ad hoc arrangements. Since the legal system currently mandates reasonable accommodations only for national examinations~\cite{xie2024noncompliant, rc-policy}, securing such accommodations in other contexts such as classrooms becomes far more challenging. Frequently, when Zhang asked for accommodations such as Braille display or test papers for a class, the response he received from his department was that \textit{``there is no prior case as a reference,''} or \textit{``we do not have budget to purchase additional equipment.''}

\BeginAccSupp{method=pdfstringdef,unicode,Alt={A scanned version of a Chinese document with multiple blurred-out personal names and details. The document is typed on white paper and discusses a student's request to take the CATTI translation qualification exam. A red official stamp appears near the bottom right, along with a date (March 30, 2021).}}
\begin{figure*}[t]
\centering
\includegraphics[width=18cm, height=10cm]{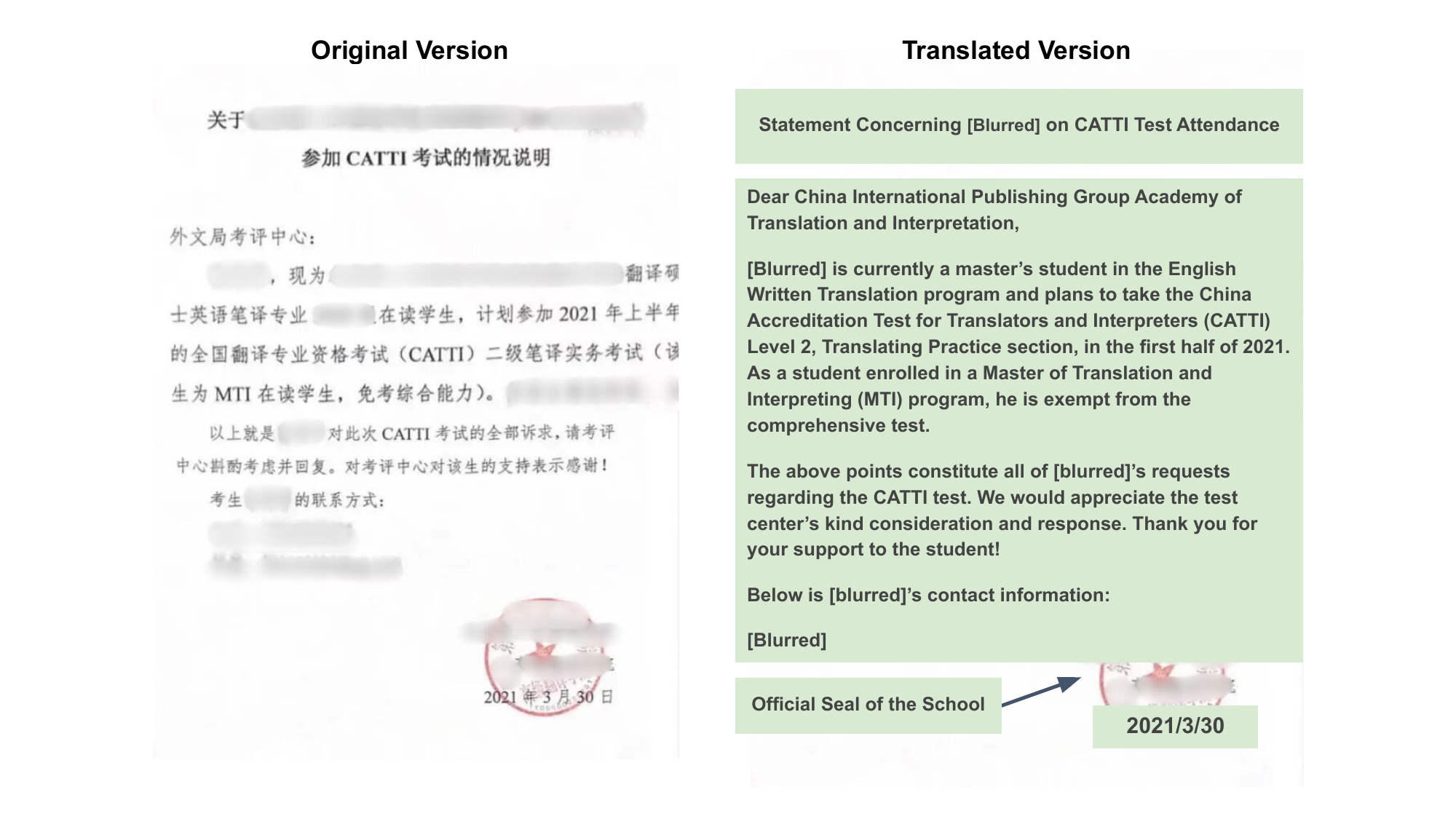}
\caption{One accommodation request letter sent by University B to the test center hosting CATTI. The letter described Zhang's condition and his request to take the test. Accommodations were not guaranteed at this point and depended on the center's response.}
\label{fig::letter}
\end{figure*}
\EndAccSupp{}

Even for national tests, where reasonable accommodations are legally mandated~\cite{rc-policy}, access to them remains uncertain due to insufficient policy guidance and limited institutional experience. For example, Zhang requested reasonable accommodations when he took the admission test and CATTI, a national qualification exam for translators and interpreters in China. In both cases, he negotiated with numerous institutions, including school departments, the local Department of Education, and other relevant offices, as the procedure was unclear to all parties (Figure \ref{fig::letter} shows an example of the many letters exchanged during these negotiations). Even with the policy~\cite{rc-policy}, these institutions often refused support, citing reasons such as, ``\textit{we do not have Braille translators,''} and ``\textit{the policy is vague, and there is no prior case.''} Meanwhile, news about changes in institutional policies continuously affected Zhang's emotions and test preparation, as these shifts signaled broader institutional attitudes toward disabled students, even when the changes did not directly apply to him. For instance, when Zhang was preparing for the admission test, ``\textit{some other university explicitly stated in its policy that it did not encourage blind or deaf students to take the test because it lacked sufficient accommodation resources such as Braille materials and sign language interpreters.}'' Faced with such uncertainty, Zhang had to continuously seek guidance from experienced individuals from his personal network, such as advocacy groups or former students who had requested accommodations. Oftentimes, he was told to \textit{``be patient and wait for certain milestones, such as registration day, before making requests to the test administrators.''}

Even when Zhang was permitted to take the tests with reasonable accommodations, the specific arrangements remained uncertain. One common issue involved compatibility with screen readers. For example, during the admission test held remotely due to COVID-19, the test paper was shared via a shared screen, which screen readers could not effectively access. In addition, anti-cheating or testing software were often incompatible with screen readers as they typically do not consider blind users. As a result, accommodations often depended on ad-hoc solutions, such as sharing test questions via WeChat (a text messaging application mostly accessible). For the CATTI exam, students who take the test were permitted to use a paper dictionary. However, the dictionary did not have a Braille version. In Zhang's case, the testing center provided a human assistant to help with look-ups. Although these accommodations provided access to the tests, they added a sense of unpredictability to the testing process. For example, in the CATTI exam, Zhang had to use NVDA instead of his preferred screen reader, as NVDA was the only accessible option in the testing environment. However, customizing NVDA and learning to use it requires considerable time and effort.

Similar experiences occurred repeatedly throughout Zhang's study at University B. Similar to findings reported in the U.S.~\cite{jain2020navigating, mack2023maintaining, shinohara2021burden, tamjeed2021understanding}, negotiating with others to secure access introduced significant labor to study. Furthermore, University B lacks centralized disability services as there were still few disabled students in the institute. As a result, the process of requesting accommodations often manifested as informal, personal, and relational work. For example, most instructors relied on PowerPoint slides for teaching. Zhang had to personally discuss with them how to make the slides accessible, and the outcome often depended largely on the teacher's personality and attitude.
\begin{quote}
    ``\textit{Some teachers agreed without hesitation. Some were willing to share but cautioned against further distribution. Some chose to read out the content on the slides during class or let other students in the class take photos of key information and send them to me. I had to read the slides using OCR (Optical Character Recognition), though.}''
\end{quote}
In some instances, instructors may refuse to offer accommodations.
\begin{quote}
    ``\textit{He refused to share the slides and would not allow photos, citing copyright concerns. He was also reluctant to provide an electronic version of the final exam paper. In desperation, I told him that if I couldn't take the exam, it would be considered a teaching failure for which he would be held responsible. Only then did he agree to have a student photograph the exam paper.}''
\end{quote}
In these instances, although requesting accommodations required effort similar to that in the U.S.~\cite{jain2020navigating, mack2023maintaining, shinohara2021burden, tamjeed2021understanding}, the process was complicated by nuanced interpersonal dynamics in a context lacking both established support policies and disability awareness.

\subsubsection{Pressure to Conform to Sighted Norms}
Another tension surfaced in our analysis was perceived pressure to conform to sighted norms. One clear piece of evidence is that the accommodations Zhang received often focused on integrating him in visually-centered tasks. An example of this is the heavy reliance on PowerPoint slides at University B.
\begin{quote}
    ``\textit{The classes I took during my graduate studies relies on visual elements a lot. Honestly, these classes were less friendly to BLV students compared to those I attended at University A. Unlike University A, most of my graduate classes were centered around PowerPoint slides. I had to speak with each instructor before the first class to discuss whether they were willing to share the slides.}''
\end{quote}
Even if instructors were willing to provide accommodations, changing the inherently visual nature of the slides proved challenging. Similarly, some traditional translation tasks were designed from sighted perspectives. For example, the admission test included a Sight Translation task, which required students to orally translate a written prompt displayed on a screen. While the institute was willing to provide accommodations, the format still favored sighted candidates:
\begin{quote}
    ``\textit{The institute allowed me to use a different testing platform that was compatible with my screen readers. However, using screen readers to read aloud the texts on the screen actually turned the task into simultaneous translation. It was more difficult than the task undertaken by sighted candidates.}''
\end{quote}
In this case, despite efforts to provide access, the task remained visual in nature, and the accommodations ultimately transformed it into something entirely different.

The pressure was further compounded by the fact that Zhang was the only blind student. It was difficult to proactively advocate for every small detail in such environments. In many instances, Zhang was careful to ensure that he did not affect others. For example, at the beginning of his study, he worried that the sound of writing on Braille paper might be disruptive.
\begin{quote}
    ``\textit{I let the class know in advance through the class group chat, saying that there might be some noise from my writing and asking for their understanding. If it affected them, they should let me know. None of my classmate raised any objections, and several showed their understanding.}''
\end{quote}
Being the only blind student also made it difficult to request accommodations when they involved infrastructural changes or coordination across multiple parties. For example, dining hall meals were a typical challenge for blind students due to the complex spatial layout and the lack of access to paper menus. In response, University A opened a separate dining area for blind students. While it is questionable whether this solution is truly inclusive, similar adjustments have been hard to implement at University B due to the lack of blind students.
\begin{quote}
``\textit{The dining hall staff [of University B] did not anticipate serving a blind student and were hesitant and unsure how to provide support. Although I offered suggestions for improvement, they remained reluctant, perceiving these proposals as relevant only to me and hesitant to make a change.}'' 
\end{quote}
In such situations, the responsibility of demonstrating the value of these changes fell on Zhang. Ultimately, he had to develop his own workarounds to navigate ongoing inaccessibility and fit into sighted-centric environments.

Pressure also arose from the pervasiveness of inaccessibility. For example, during remote teaching in the COVID period, instructors used different platforms for their teaching, including Tencent Meeting, Tencent Classroom, DingTalk, Changjiang Rain Classroom, and Xuexitong. Even when some instructors sought Zhang's input, it was difficult to determine which platform was more accessible since none of the platforms were fully accessible. As a result, Zhang did not bother advocating for specific platforms.
\begin{quote}
    ``\textit{All of these platforms presented different accessibility issues, for example, users may lose focus on certain buttons, can't send or receive text, fast-forward, pause, rewind, or read content on a shared screen. Even if some platforms were somehow usable, learning a new platform requires time. Some platforms may appear visually similar, but interactions can differ significantly for screen reader users due to inconsistent accessibility support. For instance, the same control might have a text label on platform A but none on platform B, or both may have labels but with different content.''}
\end{quote} 
Similarly, collaborative tools used for teamwork were inaccessible to screen reader users, including Tencent Docs, WPS Docs, Shimo Docs, DingTalk, and Feishu. As a result, Zhang and his teammates relied on WeChat to synchronize and update progress, even though the team still conducted most of their work on collaborative platforms. In this case, even when the team intended to include Zhang from the outset, their choice of technologies is limited by visually dominated norms, leaving them to rely on ad hoc workarounds to accommodate blind members.

\subsubsection{The Ad Hoc Nature of Change Opportunities}
On the positive side, the lack of established institutional mechanisms for accommodations could be seen as offering greater flexibility to integrate accessibility from the outset. Yet in many situations, opportunities for change were still ad hoc and depended heavily on the discretion of individual actors. For example, Zhang was allowed to report all of his needs directly to the school's leadership. However, the department was often hesitant about potential consequences in introducing new services.
\begin{quote}
    ``\textit{I mentioned to our department many times that someone from a disability advocacy organization was willing to offer free orientation and mobility training services. However, our department did not continue this conversation. They seemed to be very cautious about involving a third-party organization.}''
\end{quote} 
Oftentimes, the department considered human assistance to be the safest, most convenient, and most cost-effective form of accommodation. For example, during the COVID period, the department frequently relied on online collaborative platforms to collect students' information. Although Zhang suggested using more accessible alternatives such as Microsoft Forms, the department instead recommended having someone assist him with completing the forms. These practices, once again, imposed additional labor on disabled students, who had to work with others to navigate inaccessible systems.
\begin{quote}
   ``\textit{Human assistance became the form of support I relied on most, extending beyond filling out forms to tasks such as formatting papers and getting meals in the dining hall. Over time, I increasingly relied on the person who had become most familiar with my needs. Although we were close friends and he was willing to help, relying on him contradicted my original aim of completing tasks on my own.}'' 
\end{quote} 
Furthermore, although the school may become more aware of accessibility issues through Zhang's report, relying on Zhang to identify and manage these issues overlooks broader, more systemic opportunities for change. For example, Zhang realized that the school could have had a greater systemic impact when he took the admission test as the first blind student.
\begin{quote}
    ``\textit{In hindsight, this [making the admission test accessible for me] could have been a great opportunity to improve the test experience for students with marginalized needs, such as those with visual disabilities like me. If the needs of diverse individuals were considered from the very beginning when designing the examination process, it would significantly benefit students in the future.}''
\end{quote}
The quote above, along with other stories shared in this paper, suggests that accessibility is far from a one-time solution. It requires ongoing effort from all stakeholders, including PWD, to challenge established environments, norms, and assumptions about the world, including what is worth accessing and how access should be achieved.

\section{Discussion}
Our study extends the HCI and accessibility literature with a valuable case of a blind student's first-hand experiences of higher education in China. We now revisit the tensions identified in our analysis, considering how they are uniquely manifested in the Chinese context and their broader implications.

\subsection{The Access Dilemma in China}
Many of the issues we identified are shared across geographical areas, such as the widespread lack of accessible resources ~\cite{tang2025every, tang2024accessibility}, negotiation between multiple parties~\cite{ly2025accessibility, mack2023maintaining}, and significant labor placed on disabled students~\cite{jain2020navigating}. However, China presents its own dilemma of access: on one hand, inclusive education is increasingly valued~\cite{xie2024noncompliant} and new accessibility policies have been established~\cite{rc-policy, china-laws-a11y}. On the other hand, education for disabled students continues to be shaped by a longstanding tradition of special education and the limited development of comprehensive accessibility policies. Therefore, although reasonable accommodations are legally mandated in \textit{gaokao}, fewer than 10 students request Braille examination papers each year~\cite{hu2022opportunities}. In contrast, although the ``specialized exams and admissions'' system limits students' choice of majors, it remains the primary pathway for blind students to access higher education in China~\cite{ma2020hero}. This dilemma contrasts sharply with WEIRD contexts typically studied in HCI; for example, in the U.S., higher education institutions are technically open to all students, and approximately 78\% of BLV students attend post-secondary education after high school~\cite{newman2009post}. These regions also tend to have centralized entities providing accessibility support, such as disability offices or centers~\cite{jain2020navigating, ly2025accessibility}.

These structural conditions result in accessibility in China remaining ad hoc and shaped by ableist conceptions. As Zhang's experiences suggest, many accommodations he received depended heavily on his own self-advocacy and primarily aimed to integrate him into visual-centric environments. Even specialized education, which is intended to be accessible, is limited by ableist cultures and a systemic lack of accessible resources. While it may seem intuitive to call for new policies to improve accessibility, we recognize that addressing deeply systemic issues is not straightforward. Just as accessibility in WEIRD societies is complicated by neoliberal cultures~\cite{ly2025accessibility, apara2025}, China's traditions of elitism, intense competition in standardized testing, and limited disability awareness create their own structural barriers (see Qu's work for a more detailed analysis of structural barriers to inclusive education for disabled children in China~\cite{qu2022structural}). In light of these challenges, we draw on our findings and outline implications for research and institutions as follows.
\begin{enumerate}
    \item Aligning with previous calls~\cite{hu2022opportunities, ma2020hero}, researchers should aim to understand accessibility across diverse educational settings and explore how findings can inform actionable change. Hu identified several key directions, including: (a) qualitative studies of disabled students' experiences in both special education and mainstream schools; and (b) investigations into specific reasonable accommodations received by disabled students~\cite{hu2022opportunities}. Focusing on specific settings is crucial, as policymakers often seek concrete, localized issues within their areas of responsibility~\cite{lazar2015public, meng2021top}. Notably, a major gap remains in China regarding a basic understanding of disabled students in higher education, such as the annual number of blind students enrolling in higher education. Additionally, blind students are often grouped together with those who have low vision under the category of ``visual disabilities,'' even though they may require very different support systems~\cite{hu2022opportunities}.
    \item A case database of accommodations should be established. Research, along with Zhang's experiences, indicates that legal practice in China is challenging without prior cases for reference~\cite{xie2024noncompliant}. Building a database of relevant cases would not only serve as a starting point for legal guidance but also help gather experience in providing accommodations~\cite{yildiz2023institutional}.
    \item Practitioners and activists should work to raise awareness of the advantages and limitations of both special and mainstream education. Research suggests that in China, many blind students, their parents, and high school teachers are ignorant of disabled students' rights in \textit{gaokao}, or may discourage them from taking the test due to perceptions of lower-quality special education~\cite{ma2020hero}. While we envision a future in which higher education is accessible to all students, a crucial next step is to ensure that disabled students make informed decisions.
    \item Institutions should develop disability services and accommodation policies. There is also a need to train teachers in essential skills, such as Braille reading and writing. Crucially, our study highlights that accessibility must extend beyond providing non-visual tools to include sustained, long-term support. For instance, iterative goal-setting with mentors and regular check-ins can create space for ongoing conversations about accessibility.
    \item Ultimately, a disability network for information sharing and advocacy should be established. Zhang's experience suggests that self-advocacy remains a critical component in seeking access in China. Yet, the necessary skills and information are often inaccessible to disabled communities. Connecting stakeholders, including family members, service providers, and policymakers, is essential to developing more inclusive, rights-based services that support self-advocacy and ultimately channel resources toward advancing disability rights. Such a network is also vital for facilitating knowledge transfer across institutions, including between special education and mainstream academic institutions.
\end{enumerate}

\subsection{Exploring Access in the Shadow of Ableism}
The tensions inherent in access call for a critical research perspective that balances the long-term aspirations of disability justice with the immediate realities disabled people face in navigating ableist systems. Although our study focuses on a blind student in China, the tensions we identified are prevalent across disabilities and geographies. For example, in deaf communities, debates over signed versus oral education have persisted for decades~\cite{nakamura2006deaf, de2024illusion}. Similar tensions exist within stuttering communities, where individuals debate whether to pursue fluency or embrace stuttered speech~\cite{constantino2017rethinking}. Together, these ongoing conversations within disabled communities point to a critical question for accessibility researchers: \textit{How should we approach our work when access is both essential and problematic?}

In exploring the tensions within access, we theorize accessibility as tightly coupled with ableism, as it operates within broader ableist frameworks. Similar to Zhang's experience at the sighted-centric institute, many traditionally studied topics in accessibility are centered around visual norms, such as accessible visualizations, which are typically framed as a translation task from visuals to other modalities (see the Introduction of~\cite{zong2024umwelt}). While such frameworks highlight the unequal access to resources between disabled and non-disabled people, they may risk reinforcing ableist assumptions by framing access as a goal within existing ableist systems. As Deafblind activist John Lee Clark put in his essay \textit{Against Access}, \textit{``This is the awful function of access: to make others happy at our expense.''} ~\cite{against-access} In this provocative essay, he highlights how conventional accessibility efforts, even when well-intentioned, often prioritize access to sighted forms of information while overlooking Deafblind ways of knowing such as tactile sign languages.

Although disability justice frameworks envision futures where disability is actively affirmed, ableism is admittedly still a persistent part of disabled people's everyday realities. Although specialized education may marginalize disabled students within a broader society, it now often serves as a necessary pathway to future opportunities. Similarly, speaking fluently often signals the agency of people who stutter within ableist systems that devalue non-fluent speech, allowing them to gain privileges that might otherwise be denied~\cite{constantino2017rethinking}. As Constantino et al. noted regarding covert stuttering (the practice of hiding stuttering behaviors to appear fluent), ``\textit{Those of us who are passionate about stuttering and social change may also need to bear in mind that not everyone wants to be a radical disability.}''~\cite{constantino2022stuttering} In many cases, access to disability justice communities itself is a form of privilege~\cite{kirkham2021disability, mcdonnell2024understanding}. Given the complex tensions around access, we suggest understanding it not as merely a technical problem, but as an ongoing negotiation of agency and resistance within ableist systems. Below, we outline directions to guide research and practice toward this view.

\subsubsection{Centering Tensions in Shaping Access}
A first step toward embracing the tensions in access is to adopt more exploratory and critical research methods. Our analysis shows that access is not always a matter of clearly defined needs and solutions. Disabled students or their families often face trade-offs across different systems, who may find it hard to determine what is `better' in the lack of adequate support and under pressure to conform to mainstream norms~\cite{ma2020hero, de2024illusion}. Moreover, years of navigating imperfect accessibility systems can lower disabled individuals' expectations for accessibility services~\cite{de2021good, tang2025every} or compel them to adapt in order to survive within existing structures~\cite{ma2020hero, vogler2025barriers}. Beyond the dilemmas examined in this study, the pursuit of access could be further shaped by a wider range of intersectional values and goals across different contexts. For instance, the diversity of identities and life situations can give rise to different orientations and competing needs within disability groups~\cite{access-washing, harrington2023working}. Accessibility may also contend with competing priorities, such as organizational profitability~\cite{apara2025, ly2025accessibility}. As such, when considered in broader contexts, access is far more complex than a simple binary of ``accessible'' versus ``inaccessible.''~\cite{de2024illusion} 

To engage with these complexities, future research should move beyond solution generation and instead treat tensions in access as a generative space for surfacing issues that matter to disabled communities and for fostering ongoing dialogue and reflection, such as dilemmas in workplaces~\cite{cha2025dilemma, apara2025} and education~\cite{ma2020hero, de2024illusion}. For example, future research can engage multiple stakeholders, or center those with intersecting marginalized experiences, to uncover tensions in creating access and examine diverse perspectives on inclusion. While attending to these complexities may not immediately yield solutions, discussion around these tensions is crucial to help stakeholders make more informed decisions, and create new opportunities for collaboration and change.

\subsubsection{Including Life Course Perspectives on Access}
Our study highlights the value of incorporating a life-course perspective in understanding access, an orientation that emphasizes individuals' life transitions, trajectories, and future prospects, as well as how socioeconomic, cultural, historical, and institutional conditions shape these pathways~\cite{barros2021circumspect}. Prior research on education accessibility often focused on specific stages of schooling such as graduate schools~\cite{jain2020navigating, shinohara2021burden, shinohara2020access} or on identifying accessibility needs and issues~\cite{mack2023maintaining}. In contrast, our study reveals access as a contradictory experience that a blind student has navigated throughout their life. For blind students, including our blind researcher, access to education is not simply achieved by accommodations for specific tasks, but strategically achieved through ongoing navigation and adaptation within different systems across the lifespan. In a different yet similar context, De Meulder and Murray emphasized that inclusive education provides an ``illusion of choice'' for deaf families, highlighting the pressures that push families toward hearing education due to a gap between the desire for signing education and the lack of adequate institutional resources~\cite{de2024illusion}. To better understand how disabled people navigate access across various systems and throughout their lives, future research can draw lessons from life course studies in other fields ~\cite{mayer2009new}. For instance, comparative autoethnographies may capture richer life stories~\cite{chang2016collaborative}, and longitudinal studies involving disabled people and other stakeholders may further illuminate these trajectories~\cite{mayer2009new}.

Including a life course perspective is also important for understanding the diverse experiences of disability and how they affect people's needs and views on access. For example, disabled people may develop different orientations toward disability depending on their life experiences and family histories~\cite{nakamura2006deaf}. Additionally, desired forms of access may shift depending on a person's trajectory of disability, such as when they transition into disability~\cite{figueira2024intersecting}, and the resources available~\cite{de2024illusion}. Future work should situate individual narratives within both personal circumstances, life stories, and broader societal arrangements~\cite{sun2014being, barros2021circumspect}, when approaching access. 

\subsubsection{Exploring Radical Futures for Disabilities}
Last, we call for accessibility research to diversify its methods to expand the imagination of disability. The systematic review by Mack et al. suggests that most accessibility research relies heavily on user studies as the primary method, such as interviews, usability testing, and controlled experiments~\cite{mack2021we}. While we encourage future work to continue engaging disabled people in collectively developing design ideas such as through participatory design methods, we encourage researchers to explore more imaginative, and future-oriented methods, such as speculative design. In speculative design, speculation is often considered different from, and sometimes opposed to, direct, instrumental problem-solving. For example, Angelini et al. presented a deaf-led speculative study in which deaf participants envisioned technologies for a world inhabited exclusively by deaf signers \cite{angelini2025speculating}. Their work provided a powerful lens into what a deaf-first world might look like and the ableist assumptions embedded in existing technologies. Similarly, many radical disability movements or disability artists are envisioning disability-first worlds and deserve attention \cite{against-access, stuttering-create-time}. For example, the Protactile movement in Seattle offers a compelling vision for a tactile-centered future that challenges dominant norms and expands the scope of deafblind ways of being~\cite{against-access}. Admittedly, these imagined futures may seem radical when contrasted with the everyday realities disabled people face. However, they are powerful in exploring what disability-first futures might look like. 

\section{Conclusion}
We present an autoethnography of a blind student's higher education experience in China. By comparing specialized and mainstream education systems, we highlight the tensions involved in creating access within broader socio-political and institutional contexts. Our findings not only highlight the unique challenges and opportunities shaping educational accessibility in China but also invite critical reflections on the concept of access itself. Rather than treating access as a solution to ableism, we argue that it should be understood as an ongoing exploratory practice within ableist structures. We encourage the research community to center tensions in shaping access, incorporate life course perspectives on access, and explore radical futures for disabilities. Together, we view these three directions we propose are essential ways to support disabled individuals in exploring multiple possibilities for disability. While centering tensions and incorporating life-course perspectives can help researchers better understand the present and past of disability, engaging with radical futures expands the imagination of who disabled people need not be. This plural mindset, whether navigating or challenging existing systems, is a crucial form of disability wisdom that accessibility research should honor~\cite{constantino2017rethinking, hartblay2020disability}.

\begin{acks}
We thank the anonymous reviewers for their feedback and the crucial work of global disability activists that informed our work.
\end{acks}

\bibliographystyle{ACM-Reference-Format}
\bibliography{sample-base}
\end{document}